\documentstyle[twoside,fleqn,espcrc2,graphicx,rotate,epsfig]{article}

\newcommand{\be}{\begin{equation}}
\newcommand{\ee}{\end{equation}}

\title{ 
\vspace{-2cm}
\hfill \rm \null \hfill
 \hbox{\normalsize ADP-01-55/T487} \\
\vspace{-2mm}
 \hfill \hbox{\normalsize JLAB-THY-01-40} \\
\vspace{-1mm}
Baryon resonances from a novel fat-link fermion action
}

\author{W.~Melnitchouk\address[CSSM]{Special Research Center for the
	Subatomic Structure of Matter, and		\\
	Department of Physics and Mathematical Physics,
	Adelaide University, 5005, Australia}$^,$\address[JLab]{Jefferson Lab,
        12000 Jefferson Avenue,	Newport News, VA 23606, U.S.A.},
S.~Bilson-Thompson\addressmark[CSSM],
F.~D.~R.~Bonnet\addressmark[CSSM],
P.~D.~Coddington\addressmark[CSSM],
D.~B.~Leinweber\addressmark[CSSM],
A.~G.~Williams\addressmark[CSSM],
J.~M.~Zanotti\addressmark[CSSM],
J.~B.~Zhang\addressmark[CSSM] and
F.~X.~Lee\addressmark[JLab]$^,$\address{Center for Nuclear Studies,
	Department of Physics, \\
	The George Washington University, Washington, D.C. 20052,
	U.S.A.}}

\begin{document}

\thispagestyle{empty}

\begin{abstract}
We present first results for masses of positive and negative parity
excited baryons in lattice QCD using an ${\cal O}(a^2)$ improved gluon
action and a Fat Link Irrelevant Clover (FLIC) fermion action in which
only the irrelevant operators are constructed with fat links.  The
results are in agreement with earlier calculations of $N^*$ resonances
using improved actions and exhibit a clear mass splitting between the
nucleon and its chiral partner, even for the Wilson fermion action.
The results also indicate a splitting between the lowest $J^P={1\over
2}^-$ states for the two standard nucleon interpolating fields.
\end{abstract}

\maketitle

%%%%%%%%%%%%%%%%%%%%%%%%%%%%%%%%%%%%%%%%%%%%%%%%%%%%%%%%%%%%%%%%%%%%%%%%%%
% \section{Introduction: Why are $N^*$'s Interesting?}
\section{INTRODUCTION}

Understanding the dynamics responsible for baryon excitations provides
valuable insight into the forces which confine quarks inside baryons
and into the nature of QCD in the nonperturbative regime.  One of the
long-standing puzzles in spectroscopy has been the low mass of the
first positive parity excitation of the nucleon (the $J^P={1\over
2}^+$ $N^*(1440)$ Roper resonance) compared with the lowest lying odd
parity excitation.  Another challenge for spectroscopy is presented by
the $\Lambda^{1/2-}(1405)$, whose anomalously small mass has been
interpreted as indicating strong coupled channel effects involving
$\Sigma\pi$, $\bar K N$, $\cdots$ \cite{L1405}, and a weak overlap
with a three valence constituent quark state.

In this paper we present the first results of excited octet baryon mass
simulations using an ${\cal O}(a^2)$ improved gluon action and an
improved Fat Link Irrelevant Clover (FLIC) \cite{FATJAMES} quark
action in which only the irrelevant operators are constructed using
fat links. Configurations are generated on the new Orion computer
cluster dedicated to lattice gauge theories at the CSSM at Adelaide
University.  After reviewing in Section 2 the main
elements of lattice calculations of excited hadron masses, we describe
in Section 3 various features of interpolating fields used in this
analysis.  In Section 4 we present results for $J^P ={1\over 2}^\pm$
nucleons and hyperons.  Finally, in Section 5 we make concluding
remarks and discuss some future extensions of this work.

%%%%%%%%%%%%%%%%%%%%%%%%%%%%%%%%%%%%%%%%%%%%%%%%%%%%%%%%%%%%%%%%%%%%%%%%%%
\section{BARYONS ON THE LATTICE}

The history of excited baryons on the lattice is quite brief, although
recently there has been growing interest in finding new techniques to
isolate excited baryons, motivated partly by the experimental $N^*$
program at Jefferson Lab.  Previous work on excited baryons on the
lattice can be found in Refs.~\cite{LEIN1,LL,LEE,DWF,RICHARDS}.

Following standard notation, we define a two-point
correlation function for a baryon $B$ as:
\be
% G_B^{\alpha\alpha'}(t,{\vec p})
G_B(t,{\vec p})
\equiv \sum_{\vec{x}}\ e^{-i {\vec p} \cdot {\vec x}}
\left\langle 0 \left| T
% \chi_{B}^{\alpha}(x)\ \bar\chi_{B}^{\alpha'}(0)
\chi_B(x)\ \bar\chi_B(0)
\right| 0 \right\rangle\ ,
\ee
%
% where $\chi_B$ is a baryon interpolating field transforming positively
% under parity operation,
where $\chi_B$ is a (positive parity) baryon interpolating field,
and we have suppressed Dirac indices.
The choice of interpolating field $\chi_B$ is discussed in Section~3
below.
For large Euclidean time, the correlation function can be written as a
sum of the lowest energy positive and negative parity contributions:
\begin{eqnarray}
\label{G+-}
G_B(t,{\vec p})
&\approx& \lambda_{B^+}^2
	{ \left( \gamma \cdot p + M_{B^+} \right) \over 2 E_{B^+} }
%	e^{- E_{B^+} \, t}				\nonumber\\
	e^{- E_{B^+} \, t} \nonumber\\
% &+& { \lambda_{B^-}^2 \over 2 E_{B^-}}
 &+& \lambda_{B^-}^2
	{ \left( \gamma \cdot p - M_{B^-} \right) \over 2 E_{B^-} }
	e^{- E_{B^-} \, t}\ ,
\label{cfunc}
\end{eqnarray}
where a fixed boundary condition in the time direction is used to 
remove backward propagating states, and where
the overlap of the field $\chi_B$ with positive or negative parity
states $| B^\pm \rangle$ is parameterized by a coupling strength
$\lambda_{B^\pm}$, with
$E_{B^\pm} = \sqrt{M_{B^\pm}^2 + {\vec p\,}^2}$ the energy.
%and $u_{B^\pm}(p,s)$ a Dirac spinor.
%
The energies of the positive and negative parity states are obtained by
taking the trace of $G_B$ with the operator $\Gamma_\pm$,
where
\begin{eqnarray}
\Gamma_\pm &=&
{1\over 2} \left( 1 \pm {M_{B^\pm} \over E_{B^\pm}} \gamma_4 \right)\ .
\end{eqnarray}
For $\vec p = 0$, $E_{B^\pm} = M_{B^\pm}$ and the operator $\Gamma_\pm$
projects out the mass, $M_{B^\pm}$, of the baryon $B^\pm$.
In this case, positive parity states propagate in the 1, 1 and 2, 2
elements of the Dirac matrix of Eq.~(\ref{cfunc}), while negative parity
states propagate in the 3, 3  and 4, 4 elements.

%%%%%%%%%%%%%%%%%%%%%%%%%%%%%%%%%%%%%%%%%%%%%%%%%%%%%%%%%%%%%%%%%%%%%%%%%%
\section{INTERPOLATING FIELDS}

In this analysis we consider two types of interpolating fields which
have been used in the literature.  The notation adopted follows that
of Leinweber {\it et al}. \cite{LWD}.  For the positive parity proton
we use as interpolating fields:
\begin{eqnarray}
\label{chi1p}
\chi_1^{p +}(x)
&=& \epsilon_{abc}
\left( u^T_a(x)\ C \gamma_5\ d_b(x) \right) u_c(x)\ ,
\end{eqnarray}
and
\begin{eqnarray}
\label{chi2p}
\chi_2^{p +}(x)
&=& \epsilon_{abc}
\left( u^T_a(x)\ C\ d_b(x) \right) \gamma_5\ u_c(x)\ ,
\end{eqnarray}
where the fields $u$, $d$ are evaluated at Euclidean space-time point
$x$, $C$ is the charge conjugation matrix, $a, b$ and $c$ are color
labels, and the superscript $T$ denotes the transpose.
As pointed out by Leinweber \cite{LEIN1}, because of the Dirac
structure of the ``diquark'' in the parentheses in Eq.~(\ref{chi1p}),
the field $\chi_1^{p+}$ involves both products of {\em
upper}~$\times$~{\it upper} ~$\times$~{\it upper} and {\it
lower}~$\times$~{\it lower}~$\times$~{\it upper} components of spinors
for positive parity baryons, so that in the nonrelativistic limit
$\chi_1^{p+} = {\cal O}(1)$.  Furthermore, since the ``diquark''
couples to a total spin 0, one expects an attractive force between the
two quarks, and hence a lower energy state than for a state in which
two quarks do not couple to spin 0.

The $\chi_2^{p+}$ interpolating field, on the other hand, is known to
have little overlap with the ground state \cite{LEIN1,Bowler}.
Inspection of the structure of the Dirac matrices in Eq.~(\ref{chi2p})
reveals that it involves products of {\it upper}~$\times$~{\it
lower}~$\times$ ~{\it lower} components only for positive parity
baryons, so that $\chi_2^{p+} = {\cal O}(p^2 /E^2 )$ vanishes in the
nonrelativistic limit.  As a result of the mixing, the ``diquark''
term contains a factor $\vec \sigma \cdot \vec p$, meaning that the
quarks no longer couple to spin 0, but are in a relative $L=1$ state.
One expects therefore that two-point correlation functions constructed
from the interpolating field $\chi_2^{p+}$ are dominated by larger
mass states than those arising from $\chi_1^{p+}$.

Interpolating fields for a negative parity proton can be constructed
by multiplying the positive parity fields by $\gamma_5$, $\chi^{B-}
\equiv \gamma_5\ \chi^{B+}$, which reverses the role of the terms in
Eq.~(\ref{cfunc}).
While the masses of negative parity baryons are obtained directly from
the (positive parity) interpolating fields in (\ref{chi1p}) and
(\ref{chi2p}) by using the parity projectors $\Gamma_\pm$, it is
instructive nevertheless to examine the general properties of the
negative parity interpolating fields.
In contrast to the positive parity case, both the interpolating fields
$\chi_1^{p -}$ and $\chi_2^{p -}$ mix upper and lower components, and
consequently both $\chi_1^{p -}$ and $\chi_2^{p -}$ are ${\cal
O}(p/E)$.  Physically, two nearby $J^P = {1\over2}^-$ states are
observed in the nucleon spectrum. In simple quark models, the
splitting of these two orthogonal states is largely attributed to the
extent to which scalar diquark configurations compose the wave function
\cite{LL}.  It is reasonable to expect $\chi_1^{p-}$ to have better
overlap with scalar diquark dominated states, and thus provide a lower
effective mass in the large Euclidean time regime explored in lattice
simulations.  If the effective mass associated with the $\chi_2^p$
correlator is larger, then this would be evidence of significant
overlap of $\chi_2^{p-}$ with the higher lying $N^{{1\over2}-}$
states. In this event, further analysis directed at resolving these
two states is warranted.

%%%%%%%%%%%%%%%%%%%%%%%%%%%%%%%%%%%%%%%%%%%%%%%%%%%%%%%%%%%%%%%%%%%%%%%%%%
\section{RESULTS}

In this paper we report results of calculations of octet excited
baryon masses performed on a $16^3\times 32$ lattice at $\beta=4.60$
with a lattice spacing of $a = 0.125(2)$~fm.  The analysis is based on
a preliminary sample of 50 configurations generated on the new Orion
computer cluster at the CSSM, Adelaide.
For the gauge fields, a mean-field improved plaquette plus rectangle
action is used, while for the quark fields, the FLIC
\cite{FATJAMES} action is implemented.
In the present analysis we have imposed fixed boundary conditions in
the time direction ($U_t(\vec x, nt) = 0\ \forall\ \vec x$), and
periodic boundary conditions in spatial directions.  Although the
simulations were performed with both $n=4$ and 12 fattening sweeps,
the improved gauge fields were found to be smooth after only 4 sweeps.
Since the results with $n=4$ sweeps exhibit slightly better scaling
than those with $n=12$ \cite{FATJAMES}, we shall focus on the results
with 4 smearing sweeps.  The 12 sweep results lead to the same
conclusions as presented in the following.  Further details of the
simulations are given in Ref.~\cite{FATJAMES}.

\begin{figure}[t]
\begin{center}
\leavevmode
\epsfig{figure=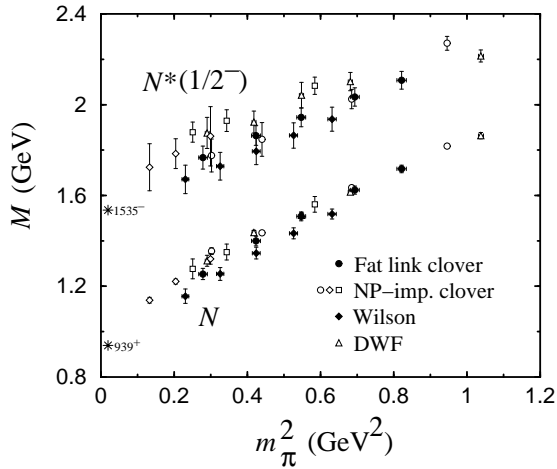,height=6.2cm}
\vspace*{-1.0cm}
\caption{Masses of the nucleon and the lowest $J^P={1\over 2}^-$
	excitation.
	The FLIC and Wilson results are
	from the present analysis, with the NP improved clover
	\protect\cite{RICHARDS} and DWF \protect\cite{DWF} results
	shown for comparison.
	The empirical $N$ and $N^*(1535)$ masses are indicated by the
	asterisks.}
\end{center}
\end{figure}

In Fig.~1 we show the $N$ and $N^*({1\over 2}^-)$ masses as a function
of the squared pseudoscalar meson mass, $m_\pi^2$.  The results of the
new simulations are indicated by the filled symbols (filled circles
are FLIC; filled diamonds are Wilson).
For comparison, we also show results from earlier simulations with
domain wall fermions (DWF) \cite{DWF} (open triangles), and a
nonperturbatively (NP) improved clover action at $\beta=6.2$
\cite{RICHARDS}.
The scatter of the different NP improved results is due to different
source smearing and volume effects: the open squares are obtained by
using fuzzed sources and local sinks, the open circles use Jacobi
smearing at both the source and sink, while the open diamonds, which
extend to smaller quark masses, are obtained from a larger lattice
($32^3 \times 64$) using Jacobi smearing.  The empirical masses of the
nucleon and the lowest ${1\over 2}^-$ excitation are indicated by the
asterisks along the ordinate.  There is excellent agreement between
the different improved actions for the nucleon mass, in particular
between the FLIC, NP improved clover \cite{RICHARDS} and DWF
\cite{DWF} results.  On the other hand, the Wilson results lie
systematically low in comparison to these due to large ${\cal O} (a)$
errors in this action \cite{FATJAMES}.
A similar pattern is repeated for the $N^*({1\over 2}^-)$ masses.
Namely, the FLIC, NP improved clover and DWF masses are in agreement
with each other, while the Wilson results again lie systematically
low.  A mass splitting of around 400~MeV is clearly visible between
the $N$ and $N^*$ for all actions, including the Wilson action,
contrary to previous claims \cite{DWF}.

\begin{figure}[t]
\begin{center}
\epsfig{figure=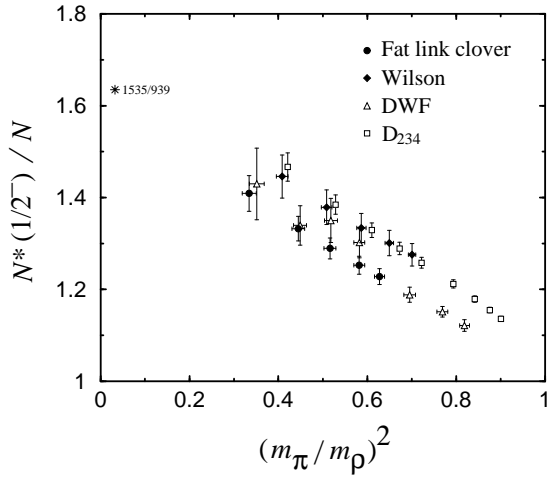,height=6.2cm}
\vspace*{-1.0cm}
\caption{Ratio of the $N^*({1\over 2}^-)$ and $N$ masses.
	The FLIC and Wilson results
	are from the present analysis, with results from the DWF
	\protect\cite{DWF} and D$_{234}$ \protect\cite{LEE} actions
	shown for comparison.
	The empirical $N^*(1535)/N$ mass ratio is denoted by the
	asterisk.}
\end{center}
\end{figure}

Figure~2 shows the ratio of the masses of the $N^*({1\over 2}^-)$ and
the nucleon.  Once again, there is good agreement between the FLIC and
DWF actions.  However, the results for the Wilson action lie above the
others, as do those for the anisotropic D$_{234}$ action \cite{LEE}.
The D$_{234}$ action has been mean-field improved, and uses an
anisotropic lattice which is relatively coarse in the spatial
direction ($a \approx 0.24$~fm).  This is an indication of the need
for nonperturbative or fat link improvement.

The mass splitting between the two lightest nearby $N^*({1\over 2}^-)$
states ($N^*(1535)$ and $N^*(1650)$) can be studied by considering the
$\chi_1$ and $\chi_2$ interpolating fields in Eqs.(\ref{chi1p}) and
(\ref{chi2p}).  Recall that the ``diquarks'' in $\chi_1$ and $\chi_2$
couple differently to spin, so that even though the correlation
functions built up from the $\chi_1$ and $\chi_2$ fields will be made
up of a mixture of many excited states, they will have dominant
overlap with different states yielding different masses
\cite{LEIN1}.
The results, shown in Fig.~3 for the FLIC action, indicate that indeed
the $N^*({1\over 2}^-)$ corresponding primarily to the $\chi_2$ field
(labeled ``$N_2^*$'') lies systematically above the $N^*({1\over
2}^-)$ associated primarily with the $\chi_1$ field (``$N_1^*$'').

\begin{figure}[t]
\begin{center}
\epsfig{figure=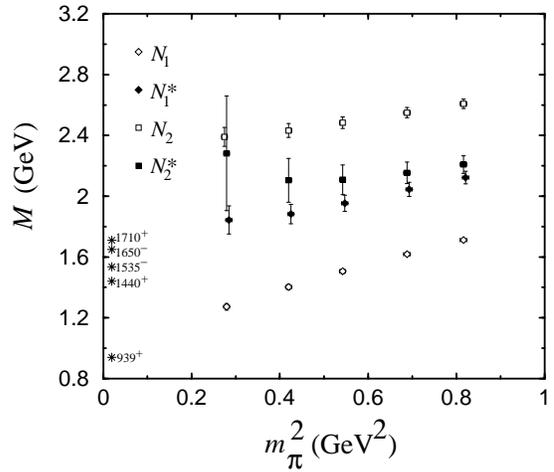,height=6.2cm}
\vspace*{-1.0cm}
\caption{Masses of the ${1\over 2}^+$ and ${1\over 2}^-$ nucleons, for the
	FLIC action.
	The positive parity ($N_1$ and $N_2$) and negative parity
	($N_1^*$ and $N_2^*$) states are constructed from the $\chi_1^p$
	and $\chi_2^p$ interpolating fields, respectively.
	The empirical masses of lowest two ${1\over 2}^\pm$ excitations
	of the nucleon are indicated by the asterisks.}
\end{center}
\end{figure}

As has long been known, the positive parity $\chi_2$ interpolating
field (``$N_2$'', which is also sometimes denoted by ``$N'({1\over
2}^+)$'') does not have good overlap with the nucleon ground state
\cite{LEIN1}, and corresponds to a state which lies around 400~MeV
above the negative parity excitation of $\chi_1$.
There is little evidence that this state is the $N^*(1440)$ Roper
resonance (first ${1\over 2}^+$ excitation of the nucleon).  While it
is possible that the Roper resonance may have a strong nonlinear
dependence on the quark mass at $m_\pi^2 < 0.2$~GeV$^2$, arising from
pion loop corrections, it is unlikely that this behavior would be so
dramatically different from that of the $N^*(1535)$ so as to reverse
the level ordering obtained from the lattice.  A more likely
explanation is that the $\chi_2$ interpolating field does not have
good overlap with either the nucleon or the $N^*(1440)$, but rather a
(combination of) excited ${1\over 2}^+$ state(s).

Recall that in a constituent quark model in a harmonic oscillator
basis the mass of the Roper is higher than the mass of the lowest
$P$-wave excitation.  The lattice data thus appear to be consistent
with the naive quark model expectation at large values of $m_q$.
Better overlap with the Roper resonance is likely to require more
exotic interpolating fields.

%%%%%%%%%%%%%%%%%%%%%%%%%%%%%%%%%%%%%%%%%%%%%%%%%%%%%%%%%%%%%%%%%%%%%%%%%%
\section{CONCLUSION}

We have presented the first results for the excited baryon spectrum
from lattice QCD using an ${\cal O}(a^2)$ improved gauge action and an
improved Fat Link Irrelevant Clover (FLIC) quark action in which only
the links of the irrelevant dimension five operators are smeared.  The
simulations have been performed on a $16^3~\times~32$ lattice at
$\beta=4.60$, providing a lattice spacing of $a = 0.125(2)$~fm.  The
analysis is based on a set of 50 configurations generated on the new
Orion computer cluster at the CSSM, Adelaide.

Good agreement is obtained between the FLIC and other
improved actions, such as the nonperturbatively improved clover 
\cite{RICHARDS} and domain wall fermion \cite{DWF} actions, for the
nucleon and its chiral partner, with a mass splitting of $\sim 400$~MeV.
Our results for the $N^*({1\over 2}^-)$ improve on those
using the D$_{234}$ \cite{LEE} and Wilson actions.
Despite strong chiral symmetry breaking, the results
with the Wilson action are still quite reasonable rendering earlier
conjectures invalid.
Using the two standard nucleon interpolating fields, we also confirm
earlier observations \cite{LL} of a mass splitting between the two
nearby ${1\over 2}^-$ states.  We find no evidence of overlap with the
${1\over 2}^+$ Roper resonance.

We have not attempted to extrapolate the lattice results to the
physical region of light quarks, since the nonanalytic behavior of
$N^*$'s near the chiral limit is not as well understood yet as that of
the nucleon \cite{MASSEXTR,ROSS}.  It is vital that future lattice
$N^*$ simulations push closer towards the chiral limit.  On a
promising note, our simulations with the 4 sweep FLIC action are able
to reach relatively low quark masses ($m_q \sim 60$--70~MeV) already.
We have also not addressed the question of to what extent quenching
may affect any of our results.  We naturally expect the effects of
quark loops to be relatively unimportant at the currently large quark
masses, although quenching may well produce some artifacts as one
nears the chiral limit.

For future work, we intend to use variational techniques to better
resolve individual excited states, for instance, those corresponding
to the $N_1^*$ and $N_2^*$ fields (using a $2\times 2$ correlator
matrix).
In order to further explore the origin of the Roper resonances, 
more exotic interpolating fields involving higher Fock states, or other
nonlocal operators should be investigated.
Finally, the present $N^*$ mass analysis will be extended in future to
include $N \to N^*$ transition form factors through the calculation of
three-point correlation functions.

%%%%%%%%%%%%%%%%%%%%%%%%%%%%%%%%%%%%%%%%%%%%%%%%%%%%%%%%%%%%%%%%%%%%%%%%%%
\vspace*{0.3cm}
We thank D.G.~Richards for providing the data points from
Ref.~\cite{RICHARDS}.  This work was supported by the Australian
Research Council, and the U.S. Department of Energy contract
\mbox{DE-AC05-84ER40150}, under which the Southeastern Universities
Research Association (SURA) operates the Thomas Jefferson National
Accelerator Facility (Jefferson Lab).

%%%%%%%%%%%%%%%%%%%%%%%%%%%%%%%%%%%%%%%%%%%%%%%%%%%%%%%%%%%%%%%%%%%%%%%%%%

%%%%%%%%%%%%%%%%%%%%%%%%%%%%%%%%%%%%%%%%%%%%%%%%%%%%%%%%%%%%%%%%%%%%%%%%%%


\begin{thebibliography}{40}

\bibitem{L1405}
R.H.~Dalitz and J.G.~McGinley,
Oxford U. preprint TP-47-80 (1980);
%
E.A.~Veit, B.K.~Jennings, R.C.~Barrett and A.W.~Thomas,
Phys. Lett. B {\bf 137} (1984) 415;
%
P.B.~Siegel and W.~Weise,
Phys. Rev. C {\bf 38} (1988) 2221.

\bibitem{FATJAMES}
J.M.~Zanotti {\it et al}.,
hep-lat/0110216;
and these proceedings.

\bibitem{LEIN1}
D.B.~Leinweber,
Phys. Rev. D {\bf 51} (1995) 6383.

\bibitem{LL}
F.X.~Lee and D.B.~Leinweber,
Nucl. Phys. Proc. Suppl. {\bf 73} (1999) 258.

\bibitem{LEE}
F.X.~Lee,
Nucl. Phys. Proc. Suppl. {\bf 94} (2001) 251.

\bibitem{DWF}
S.~Sasaki, T.~Blum and S.~Ohta,
hep-lat/0102010.

\bibitem{RICHARDS}
D.G.~Richards,
Nucl. Phys. Proc. Suppl. {\bf 94} (2001) 269;
G\"ockeler {\it et al}.,
hep-lat/0106022.

\bibitem{LWD}
D.B.Leinweber, R.W.Woloshyn and T.Draper,
Phys. Rev. D {\bf 43} (1991) 1659.

\bibitem{Bowler}
K.~Bowler {\it et al}., Nucl. Phys. {\bf B240} (1984) 213.

\bibitem{MASSEXTR}
D.B.~Leinweber, A.W.~Thomas, K.~Tsushima and S.V.~Wright,
Phys. Rev. D {\bf 61} (2000) 074502.

\bibitem{ROSS}
R.D.~Young {\it et al.}, 
hep-lat/0111041; and these proceedings.
 
\end{thebibliography}
\end{document}